\documentclass[aps,prd, nofootinbib,preprintnumbers]{revtex4}%
\usepackage{amsmath,amsfonts,amssymb,graphicx,graphics,color, hyperref}
\usepackage[latin9]{inputenc}
\usepackage{pifont}
\usepackage{natbib}
\usepackage{subfigure,epsfig,epstopdf}
\usepackage{bm}
\usepackage{enumerate}
\usepackage{amsmath}
\usepackage{amsfonts}
\usepackage{amssymb}
\usepackage{graphicx}%
\providecommand{\U}[1]{\protect\rule{.1in}{.1in}}

\newcommand{\be}{\begin{equation}}
\newcommand{\ee}{\end{equation}}
\newcommand{\bea}{\begin{eqnarray}}
\newcommand{\eea}{\end{eqnarray}}

\begin{document}
\title{Magnetized Baryonic layer and a novel BPS bound in the
gauged-Non-Linear-Sigma-Model-Maxwell theory in (3+1)-dimensions \\through Hamilton-Jacobi equation}
\author{Fabrizio Canfora$^{1,2}$}
\author{}
\affiliation{$^{1}$Facultad de Ingenieria, Arquitectura y Dise$\widetilde{n}$o, Universidad
San Sebastian, sede Valdivia, General Lagos 1163, Valdivia 5110693, Chile.}
\affiliation{$^{2}$Centro de Estudios Cient\'ificos (CECs), Avenida Arturo Prat 514,
Valdivia, Chile}
\email{fabrizio.canfora@uss.cl}

\begin{abstract}
It is show that one can derive a novel BPS bound for the gauged
Non-Linear-Sigma-Model (NLSM) Maxwell theory in (3+1) dimensions which can
actually be saturated. Such novel bound is constructed using Hamilton-Jacobi
equation from classical mechanics. The configurations saturating the bound
represent Hadronic layers possessing both Baryonic charge and magnetic flux.
However, unlike what happens in the more common situations, the topological
charge which appears naturally in the BPS bound is a non-linear function of
the Baryonic charge. This BPS bound can be saturated when the surface area of
the layer is quantized. The far-reaching implications of these results are
discussed. In particular, we determine the exact relation between the magnetic
flux and the Baryonic charge as well as the critical value of the Baryonic
chemical potential beyond which these configurations become thermodynamically unstable.

\end{abstract}
\maketitle
\tableofcontents

\section{Introduction}

The phase diagram of the low energy limit of QCD under extreme conditions
(like finite density, low temperatures and when strong magnetic fields are
involved) is a very hard nut to crack (see \cite{R0} \cite{R11} \cite{R2} and
references therein). Not only analytic perturbative methods fail, but also
Lattice QCD (LQCD henceforth) is not very effective both due to the sign
problem as well as due to the magnetic field (see \cite{sign1} \cite{sign2}
\cite{sign3m} \cite{HIC}\ \cite{QGBook} and references therein). In the
Infra-Red (IR henceforth) the analysis of the interplay between strong
interactions and electromagnetic fields still exhibits difficult unsolved
problems. In IR phase, the main role is played by the topological solitons of
QCD (see \cite{[4]} \cite{[5]} \cite{[6]} \cite{R11} and references therein).
Topological solitons are characterized by a non-vanishing topological charge
which prevents such classical configurations from decaying into the trivial
vacuum. There are many relevant physical effects which are genuine features of
the fact that a finite amount of topological charge is "forced to live" within
a finite spatial region. One of the most important is the appearance of
non-homogeneous Baryonic condensates (which would not form in free space). In
(1+1)-dimensional models (where the
\"{}%
solitons%
\"{}
are kinks, such as in the Gross-Neveu model and its variants: see, for
instance, \cite{[8]}) it has been possible to show, thanks to integrability,
that there is a finite region in the phase diagram (which appears at finite
density) where kinks crystals (namely, ordered arrays of kinks) dominate
(depending on the value of the isospin chemical potentials). Similar results
have been obtained in higher dimensional models (under the assumption that the
main fields only depend on one spatial coordinate: see \cite{[9]} \cite{[10]}
and references therein). These results strongly suggest that in the low energy
limit of QCD there should appear non-homogeneous condensates similar to the
ones appearing in superconductors \cite{LOFF1} \cite{LOFF2}. Indeed, in (3+1)
dimensions there are strong numerical as well as phenomenological evidences
(see \cite{pasta1}, \cite{pasta2}, \cite{pasta3}, \cite{pasta4} \cite{[12]}
and references therein) suggesting that non-homogeneous Baryonic condensates
do appear at low energies and temperatures when a finite amount of Baryonic
charge is forced into a finite spatial volume. The nice regular shapes of
these condensates lead to the name \textit{nuclear pasta phases}. In the
references mentioned above, it has been shown that at finite Baryon densities
\textquotedblleft hadronic tubes\textquotedblright\ (\textit{nuclear
spaghetti}), "Hadronic layers" (\textit{nuclear lasagna}) and so on do appear.
The numerical simulations of these systems are very challenging and the
situation becomes much worse when the self-consistent electromagnetic
interactions of these non-homogeneous Baryonic condensates are taken into
account. Consequently, the development of novel analytic tools able to support
the numerical simulations is a priority. In the present paper we will analyze
the prototype of strongly interacting configurations where the techniques used
in \cite{[8]} \cite{[9]} \cite{[10]} are not especially effective due to the
lack of integrability: magnetized non-homogeneous Baryonic condensates in
(3+1) dimensions where the fields necessarily depend on all the spatial coordinates.

We will consider here the (3+1)-dimensional gauged non-linear sigma model
(gauged NLSM henceforth) Maxwell theory in the case of $SU(2)$ isospin global
symmetry which is one of the most relevant effective field theories: see
\cite{[4]} \cite{BaMa} and references therein. One of the main reasons is that
such theory is the basic building block of Chiral Perturbation Theory (CPT
henceforth, see \ \cite{CPT1} \cite{CPT2} \cite{CPT3} \cite{1m} \cite{2m}
\cite{3m} \cite{4m} and references therein) being closely related to the low
energy limit of QCD in (3+1) dimensions). Moreover, the well known arguments
clarifying the interpretation of the topological charge as Baryonic charge
(see \cite{Skyrme1} \cite{Skyrme2} \cite{Skyrme3} \cite{Skyrme4} and
references therein) do not actually need the explicit presence of the Skyrme
term in the gauged NLSM but rather of a mechanism to stabilize the solitons
(so that the search for stabilizing mechanism in the gauged NLSM Maxwell
theory acquires high priority). The analysis of the conditions allowing the
appearance of topologically stable non-homogeneous condensates, besides its
intrinsic interest, has important consequences for the thermodynamics of dense
nuclear/quark matter (see \cite{R0} \cite{R2} and references therein).
Moreover, when the $SU(2)$ valued scalar field is assumed to be homogeneous
and the corresponding spatial fluctuations are neglected, many important
physical effects are missed (see \cite{ex4d6} \cite{ex4d4} and references
therein). The more common techniques (which allow both in the Yang-Mills-Higgs
and in the Abelian-Higgs models to describe stable topological solitons such
as instantons, monopoles and vortices) do not work in the case of the gauged
NLSM Maxwell theory. The reason is that, in the case in which the minimal
coupling with electromagnetism is not taken into account, there is no obvious
BPS bound to saturate for static configurations due to the Derrick scaling
argument. On the other hand, when the minimal coupling is turned on, there is
some hope (as in the Abelian-Higgs model at critical coupling one can find
topologically non-trivial vortices, which would not appear in the spectrum in
the absence of gauge field). Unfortunately, no explicit example of BPS bound
which can be saturated in the (3+1)-dimensional gauged NLSM Maxwell theory has
been found so far. This the main reason why many authors proposed
modifications to the NLSM and Skyrme theories (see \cite{gaugsol5}
\cite{gaugsol6} \cite{gaugsol7} and references therein): the idea is to modify
the theory in order to allow the appearance of BPS bound which can actually be
saturated. However, the most important question remains open:

\textit{Is it possible to find novel BPS bounds in the gauged NLSM Maxwell
theory (without any modification) in terms of different topological charges
which can actually be saturated?}

It is worth emphasizing that the obvious candidate to be the topological
charge appearing in the BPS bound (which is the Baryonic charge) does not work
very well since the available numerical solutions are always above the bound
(see \cite{[4]} \cite{[5]} \cite{[6]} \cite{R11} and references therein). In
the present paper we will construct non-homogeneous magnetized Hadronic
condensates depending on all the spatial coordinates in such a way to have
non-vanishing Baryonic charge and magnetic flux. This approach also helps to
avoid the Landau-Peierls\footnote{They showed that in an isotropic system in
three or fewer dimensions, thermal fluctuations destroy condensates depending
on only one spatial coordinate.} \cite{ex4d6} (which cannot be applied when
the fields depend in a non-trivial way on the three spatial coordinates).
Secondly, a magnetized topologically non-trivial condensates cannot decay into
a condensate with vanishing topological charge and magnetic flux. It is worth
emphasizing that magnetize Baryonic layers are extremely relevant in many
situations such as heavy ions collisions and neutron stars (see \cite{sign1}
\cite{sign2} \cite{sign3m} \cite{HIC}\ \cite{QGBook} \cite{[12]} and
references therein).

The techniques introduced in \cite{canfora2}, \cite{56}, \cite{56b},
\cite{58}, \cite{58b}, \cite{ACZ}, \cite{CanTalSk1}, \cite{canfora10},
\cite{Fab1}, \cite{gaugsk}, \cite{Canfora:2018clt} \cite{lastEPJC}
\cite{crystal1} allowed the construction of analytical solutions that describe
multi-solitons at finite density for both, in the non-linear sigma model as
well as in the Skyrme model minimally coupled to the Maxwell theory. However,
in this approach the electromagnetic field self-consistently generated by the
non-homogeneous Baryonic condensates has necessarily the electric components
of the same order as the magnetic components. Therefore, in all the situations
in which the magnetic field dominates, this framework must be modified. This
is the aim of the present paper in which the theory of Hamilton-Jacobi
equation will be used to derive a novel BPS bound which can be saturated for
magnetized configurations.

This paper is organized as follows: in the second section the gauged
NLSM-Maxwell theory will be introduced. In the third section the ansatz to
describe magnetized Baryonic layers will be described. In the fourth section
the technique to derive novel BPS bound using the Hamilton-Jacobi equation in
classical mechanics will be explained. In the fifth section such technique
will be applied to construct analytically BPS magnetized Baryonic
configurations. In the sixth section, the main physical characteristics of
these non-homogeneous condensates will be presented. In the final section some
conclusions will be drawn.

\section{The gauged non-linear sigma model}

At finite density, a fundamental question which has been only partially
addressed is whether or not the gauged non-linear sigma model admits
topologically non-trivial non-homogeneous condensates with crystal-like
structures. The Chiral Soliton Lattice (CSL henceforth: see \cite{ex4d5}
\cite{ex4d1} \cite{ex4d2} \cite{ex4d3} and references therein) is a very
interesting (3+1)-dimensional example in which the Pionic field only depends
on one spatial coordinate. This model can be supported by strong external
fields (as explained in the above references). However, the question of the
magnetic field generated self-consistently by topologically non-trivial
non-homogeneous condensates depending in a non-trivial way on all the spatial
coordinates remains open. One would like to know the magnetic field generated
by these condensates, whether or not such magnetic field grows with the
Baryonic charge and so on. With the available numerical methods, to answer to
such a question cannot be achieved (yet). In the present paper we will
construct non-homogeneous condensates depending on all the spatial coordinates
in such a way to have non-vanishing topological charge. Therefore, the
classical no-go argument by Landau and Peierls \cite{ex4d6} is avoided
directly and, moreover, the issue of the stability (which, usually, is a quite
difficult problem to analyze) in the present case will be clarified by the
novel BPS bound which will be derived in the next sections.

As it can be shown using, for instance, Chiral Perturbation Theory ($\chi$PT
henceforth, see \cite{1m} \cite{2m} \cite{3m} \cite{4m}), the low energy QCD
action\footnote{We have chosen the units in such a way that both the speed of
light and the Planck divided by $2\pi$ are set to 1: $c=1$, $\hslash=1$.} (to
order $\mathit{O}(p^{2})$) reads
\begin{equation}
S=\int\frac{d^{4}v}{4}\left\{  -\widehat{K}\ \mathrm{Tr}\left[  \Sigma^{\mu
}\Sigma_{\mu}\right]  -\frac{1}{e^{2}}F_{\mu\nu}F^{\mu\nu}\right\}  \,,
\label{action0}%
\end{equation}
where $e$ is the electric charge and $\widehat{K}$ is the coupling constant of
the gauged NLSM. The constant $\widehat{K}$ is related with the Pions decay
constant $f_{\pi}$ as follows (see \cite{Skyrme4}):
\begin{equation}
\widehat{K}=\frac{\left(  f_{\pi}\right)  ^{2}}{4}\ ,\ \ \ f_{\pi}%
\approx130\ \ MeV\ . \label{action0.1}%
\end{equation}
The above action in Eq. (\ref{action0}) can be rewritten as follows:
\begin{equation}
S=\frac{1}{4e^{2}}\int\frac{d^{4}v}{4}\left\{  -K\ \mathrm{Tr}\left[
\Sigma^{\mu}\Sigma_{\mu}\right]  -F_{\mu\nu}F^{\mu\nu}\right\}  \,,\ \ K=e^{2}%
\widehat{K}\ . \label{action1}%
\end{equation}
The above form of the action is very convenient since the field equations will
only depend on the "rescaled" Pions decay constant $K$ defined here above
while the factor $1/e^{2}$ will enter only as global factor in the definition
of the energy density and total energy. The pions mass could also be included,
however in the present case can be safely neglected as the mass of the
solitons which will be constructed here is much bigger than the Pions mass.
Moreover
\[
\Sigma_{\mu}=\Sigma^{-1}D_{\mu}\Sigma=\Sigma_{\mu}^{j}t_{j},\ F_{\mu\nu
}=\partial_{\mu}A_{\nu}-\partial_{\nu}A_{\mu}\,,\ t_{j}=i\sigma_{j}\,,
\]
where $\sigma_{i}$ are the Pauli matrices and the $SU(2)$ valued $\Sigma$
field can be parametrized using Euler angles (as any element of $SU(2)$ can be
written in a unique way in Euler parametrization):
\begin{equation}
\Sigma=\exp\left(  t_{3}G_{1}\right)  \exp\left(  t_{2}G_{2}\right)
\exp\left(  t_{3}G_{3}\right)  \,, \label{eq:defSigma}%
\end{equation}
where $G_{j}\left(  x^{\mu}\right)  $ (with $j=1,2,3$) are the three scalar
degrees of freedom of $SU(2)$. The covariant derivative is defined as
\begin{equation}
\ D_{\mu}\Sigma=\partial_{\mu}\Sigma+A_{\mu}\left[  t_{3},\Sigma\right]  \ ,
\label{sky2.75}%
\end{equation}
where $\partial_{\mu}$\ is the usual partial derivative. The field equations
for the gauged NLSM read
\begin{equation}
D_{\mu}\Sigma^{\mu}=0\ , \label{eq:NLSM}%
\end{equation}%
\begin{equation}
\partial_{\mu}F^{\mu\nu}=J^{\nu}\ , \label{eq:maxwellNLSM}%
\end{equation}
where the current $J^{\mu}$ is given by
\begin{equation}
J^{\mu}=\frac{K}{2}\text{Tr}\left[  \widehat{O}\Sigma^{\mu}\right]
\ ,\qquad\widehat{O}=\Sigma^{-1}t_{3}\Sigma-t_{3}\ . \label{maxcurrent}%
\end{equation}
The energy-momentum tensor is given by
\[
T_{\mu\nu}=\frac{1}{e^{2}}\left\{  -\frac{K}{2}\mathrm{Tr}\left[  \Sigma_{\mu
}\Sigma_{\nu}-\frac{1}{2}g_{\mu\nu}\Sigma^{\alpha}\Sigma_{\alpha}\right]
+\ \bar{T}_{\mu\nu}\right\}  \,,
\]
with
\begin{equation}
\bar{T}_{\mu\nu}=-F_{\mu\alpha}F_{\nu}^{\;\alpha}+\frac{1}{4}F_{\alpha\beta
}F^{\alpha\beta}g_{\mu\nu}\ . \label{tmunu(1)}%
\end{equation}

\section{Magnetized BPS Baryonic layers}

As it happens with the superconductors described by the Abelian-Higgs model,
the physical properties are largely determined by the magnetic field. This is
the reason why it is so important to analyze the physics of magnetized
Baryonic layers. Unfortunately, in this situation, the techniques developed in
\cite{canfora2}, \cite{56}, \cite{56b}, \cite{58}, \cite{58b}, \cite{ACZ},
\cite{CanTalSk1}, \cite{canfora10}, \cite{Fab1}, \cite{gaugsk},
\cite{Canfora:2018clt} \cite{lastEPJC} cannot be applied as these require the
presence also of an electric field. Therefore, in order to answer to the
question "how can we describe Baryonic layers generating in a self-consistent
way only a magnetic field without electric component", we need a different
approach. In other words, we need to solve the coupled system of three
non-linear PDEs arising from the gauged NLSM minimally coupled to a $U(1)$
gauge potential together with the corresponding four Maxwell equations (where
the $U(1)$ current arising from the NLSM in Eq.(\ref{maxcurrent})) in the case
in which the layers generate a pure magnetic field (no electric components) to
be as close as possible to the solitonic solutions of the Abelian-Higgs model.
Since the main physical motivation of the present work is to study the
appearance of topologically non-trivial non-homogeneous condensates at finite
density, finite volume effects are of crucial importance. The easiest way to
take them into account is to use the following flat metric
\begin{equation}
ds^{2}=-dt^{2}+L_{r}^{2}dr^{2}+L^{2}\left(  dx^{2}+dy^{2}\right)  \ ,
\label{Minkowski}%
\end{equation}
where $4\pi^{3}L_{r}L^{2}$ is the volume of the box in which the gauged
solitons are living. The adimensional Cartesian coordinates $r$, $x$ and $y$
have the ranges
\begin{equation}
0\leq r\leq2\pi\ ,\quad0\leq x\leq\pi\ ,\quad0\leq y\leq2\pi\ .
\label{period0}%
\end{equation}
As it will be clear when analyzing the magnetized Baryonic layers in the next
sections, the coordinates $x$ and $y$ are coordinates tangent to the layers
while the coordinate $r$ is orthogonal to the layers. The reason is that such
layers are localized in the $r$ direction as the corresponding energy and
Baryon densities only depend on $r$. Thus, (once $r$ is fixed at the position
of the layer) one moves along the layers by moving $x$ and $y$. On the other
hand, one moves away from the layer by moving $r$ away from the position of
the layer (which is the value of $r$ at which the energy-density has its
maximum). Consequently, the area $A$ of the layer is
\begin{equation}
A=2\pi^{2}L^{2}\ . \label{area1}%
\end{equation}
Following the strategy of \cite{Fab1}, \cite{gaugsk} and
\cite{Canfora:2018clt}, let us consider the following static ansatz for the
$SU(2)$ valued scalar field and for the $U(1)$ gauge field
\begin{align}
\Sigma(x^{\mu})  &  =\exp\left(  py\ t_{3}\right)  \exp\left(  H(r)\ t_{2}%
\right)  \exp\left(  px\ t_{3}\right)  \ ,\label{ansatzmagnetico1}\\
A_{\mu}  &  =\left(  0,0,\frac{p}{2}-u\left(  r\right)  ,-\frac{p}{2}+u\left(
r\right)  \right)  \ ,\ p\neq0\ ,\ p\in%
\mathbb{N}%
\nonumber
\end{align}
where $p$ must be an integer according to the theory of Euler angles for
$SU(N)$ (see, for instance, \cite{Euler1} \cite{Euler2} \cite{Euler3}\ and
references therein). Thus, in the above ansatz (which depend in a non-trivial
way on the three spatial coordinates), only two scalar degrees of freedom
(namely, the two profiles $H(r)$ and $u(r)$) are turned on (in a similar way
as it happens with the vortex in the superconductors where the field equations
reduce to two coupled equations for the Higgs profile and for one component of
the $U(1)$ gauge potential). The first relevant observation is that the full
system of seven coupled non-linear field equations is reduced to the following
two coupled ODEs:%
\begin{align}
H^{\prime\prime}+4\left(  \frac{L_{r}}{L}\right)  ^{2}\sin\left(  2H\right)
\left(  \left(  \frac{p}{2}\right)  ^{2}-u^{2}\right)   &  =0\label{equ1}\\
u^{\prime\prime}-4KL_{r}^{2}\sin^{2}\left(  H\right)  u  &  =0\ , \label{equ2}%
\end{align}
where, of course, $\ $%
\[
X^{\prime}=\frac{d}{dr}X\ ,\ X^{\prime\prime}=\frac{d^{2}}{dr^{2}}X\
\]
and so on. Moreover, with the above ansatz, the energy-density $T_{00}$
reduces to%
\begin{equation}
T_{00}=\frac{1}{e^{2}}\left\{  \frac{K}{L^{2}}\left[  p^{2}\cos^{2}\left(
H\right)  +4\sin^{2}\left(  H\right)  u^{2}\right]  +\frac{K\left(  H^{\prime
}\right)  ^{2}}{2L_{r}^{2}}+\frac{\left(  u^{\prime}\right)  ^{2}}{\left(
L_{r}L\right)  ^{2}}\right\}  \ , \label{energydensity1}%
\end{equation}
and the field equations corresponding to the ansatz in Eq.
(\ref{ansatzmagnetico1}) can be derived using the above energy-density as
starting point for a variational principle.

The topological (Baryonic) charge of the gauged non-linear sigma
model~\cite{Skyrme1} \cite{Skyrme2} \cite{Skyrme3} \cite{Skyrme4} \cite{Bala1}
\cite{Witten,gaugesky1} is given by
\begin{equation}
B=\frac{1}{24\pi^{2}}\int_{S}\rho_{B}\ ,\ \ \rho_{B}=\rho_{B1}+\rho
_{B2}\ ,\label{new4.1}%
\end{equation}
where $S$ is the three-dimensional $t=$constant hypersurface defined by the
metric in Eqs. (\ref{Minkowski}) and (\ref{period0}) while
\begin{align}
\rho_{B1} &  =\epsilon^{ijk}\text{Tr}\left\{  \left(  \Sigma^{-1}\partial
_{i}\Sigma\right)  \left(  \Sigma^{-1}\partial_{j}\Sigma\right)  \left(
\Sigma^{-1}\partial_{k}\Sigma\right)  \right\}  \ ,\label{rhoSk}\\
\rho_{B2} &  =-3\epsilon^{ijk}\text{Tr}\left\{  \partial_{i}\left[  A_{j}%
t_{3}\left(  \Sigma^{-1}\partial_{k}\Sigma+\left(  \partial_{k}\Sigma\right)
\Sigma^{-1}\right)  \right]  \right\}  \,,\label{rhoMax}%
\end{align}
are the two topological density contributions ($A_{j}$ being the spatial
components of the gauge potential). With the above ansatz we get%
\begin{equation}
\rho_{B}=J_{0}^{B}=-12p\left[  u(1+\cos(2H))\right]  ^{\prime}\ .\label{BCh1}%
\end{equation}

\section{The BPS bound: a novel strategy based on the Hamilton-Jacobi
equation}

Before going back to the case of interest in this paper, let us outline a
general strategy to derive a novel BPS bound using Hamilton-Jacobi equation.
To the best of author's knowledge, this strategy to construct BPS bounds is
new. Suppose one has the positive definite energy-density $T_{00}$ of a static
configurations of two (or more) interacting degrees of freedom, $J_{1}$ and
$J_{2}$ (in the present case, $J_{1}$ and $J_{2}$ are related to $H$ and $u$)
which depend on one spatial coordinate (let%
\'{}%
s say, $r$):%
\begin{equation}
T_{00}=\frac{\left(  \partial_{r}J_{1}\right)  ^{2}}{2}+\frac{\left(
\partial_{r}J_{2}\right)  ^{2}}{2}+V\left(  J_{1},J_{2}\right)  \ ,
\label{prova}%
\end{equation}
where $V\left(  J_{1},J_{2}\right)  $ is the interaction term between the two
degrees of freedom (which is supposed to be known). Being the configuration
static (by hypothesis) the field equations for $J_{1}$ and $J_{2}$ can be
derived using $T_{00}$ as action density. In order to get a BPS bound for
$J_{1}$ and $J_{2}$ we would like to sum two quantities (let us call them
$\Gamma_{1}$ and $\Gamma_{2}$) to the gradients of $J_{1}$ and $J_{2}$ with
the following properties:%
\[
\frac{\left(  \partial_{r}J_{1}\pm\Gamma_{1}\right)  ^{2}}{2}+\frac{\left(
\partial_{r}J_{2}\pm\Gamma_{2}\right)  ^{2}}{2}=T_{00}+total\ derivative\ .
\]
The above requirement implies that%
\begin{align}
\frac{\left(  \Gamma_{1}\right)  ^{2}}{2}+\frac{\left(  \Gamma_{2}\right)
^{2}}{2}  &  =V\left(  J_{1},J_{2}\right)  \ ,\label{cond1}\\
\Gamma_{1}\partial_{r}J_{1}+\Gamma_{2}\partial_{r}J_{2}  &
=total\ derivative\ . \label{cond2}%
\end{align}
Moreover, Eq. (\ref{cond2}) can be satisfied by requiring that%
\begin{equation}
\Gamma_{1}=\frac{\partial W}{\partial J_{1}}\ ,\ \Gamma_{2}=\frac{\partial
W}{\partial J_{2}}\ ,\ \ W=W(J_{1},J_{2})\ , \label{cond2bis}%
\end{equation}
since (if the above condition is satisfied) then%
\[
\Gamma_{1}\partial_{r}J_{1}+\Gamma_{2}\partial_{r}J_{2}=\partial_{r}\left(
W(J_{1},J_{2})\right)  \ .
\]
Hence, putting together Eqs. (\ref{cond2bis}) and (\ref{cond1}) we arrive at
the main requirement able to provide a BPS bound:%
\begin{align}
\left(  \frac{\partial W}{\partial J_{1}}\right)  ^{2}+\left(  \frac{\partial
W}{\partial J_{2}}\right)  ^{2}  &  =2V\left(  J_{1},J_{2}\right)
\ \Rightarrow\label{condBPSfinal}\\
total\ derivative  &  =\partial_{r}W\ .\nonumber
\end{align}
In other words, if one is able to solve the above Hamilton-Jacobi equation Eq.
(\ref{condBPSfinal}) for the function $W(J_{1},J_{2})$ where the role of the
potential is played by $V\left(  J_{1},J_{2}\right)  $, then one get the
following bound:%
\[
E\geq\left\vert W\left(  2\pi\right)  -W\left(  0\right)  \right\vert \ .
\]
The corresponding BPS equation are%
\[
\partial_{r}J_{k}\pm\frac{\partial W}{\partial J_{k}}=0\ ,\ \ k=1,2\ .
\]
Due to the huge amount of literature on the Hamilton-Jacobi equation, this
strategy is extremely convenient and, as we will now show, it allows to derive
new topological bound where the corresponding topological charge is a
non-linear function of the "obvious topological charge" which one would
consider, at a first glance, as candidate to derive BPS equations. Although
the relations between the Hamilton-Jacobi equation and SUSY are well known
(see \cite{SUSYHJ1} \cite{SUSYHJ2}\ and references therein), such approach has
not been used so far to derive novel BPS bounds in (3+1)-dimensional field
theories which are not supersymmetric (at least, not in the usual sense) such
as the (3+1)-dimensional gauged NLSM Maxwell theory.

\section{Application: magnetized BPS Baryonic layers}

A beautiful application of the above strategy is the derivation of a BPS bound
for magnetized Baryonic layers. As it has been already mentioned, it is not
possible to solve the system in Eqs. (\ref{equ1}) and (\ref{equ2}) using the
techniques in \cite{canfora2}, \cite{56}, \cite{56b}, \cite{58}, \cite{58b},
\cite{ACZ}, \cite{CanTalSk1}, \cite{canfora10}, \cite{Fab1}, \cite{gaugsk},
\cite{Canfora:2018clt} \cite{lastEPJC}. The reason is that in those
references, the non-homogeneous Baryonic condensates produce electromagnetic
field in which the electric component is of the same order as the magnetic
one. In those cases, the complete set of field equations can be solved
analytically since the electromagnetic field is of force-free type. In the
case of non-homogeneous condensates generating pure magnetic fields (as
vortices in superconductors) one must use different techniques. Despite the
fact that in the (3+1)-dimensional gauged NLSM - Maxwell theory, usually the
search for BPS bound has not produced explicit analytic solutions in
topologically non-trivial sectors, the technique introduced here above open a
new unexpected window.

Using the ideas described in the previous section together with the
definitions%
\[
\Gamma_{1}=\frac{L_{r}}{K^{1/2}}\frac{\partial W}{\partial H}\ ,\ \ \Gamma
_{2}=\frac{L_{r}L}{\sqrt{2}}\frac{\partial W}{\partial u}\ ,
\]
one obtains the following Hamilton-Jacobi equation associated to the
energy-density in Eq. (\ref{energydensity1}):%
\begin{equation}
\frac{L_{r}^{2}}{2K}\left(  \frac{\partial W}{\partial H}\right)  ^{2}+\left(
\frac{L_{r}L}{2}\right)  ^{2}\left(  \frac{\partial W}{\partial u}\right)
^{2}=\frac{K}{L^{2}}\left[  p^{2}\cos^{2}\left(  H\right)  +4\sin^{2}\left(
H\right)  u^{2}\right]  \ . \label{HJBPS1}%
\end{equation}
Such Hamilton-Jacobi equation can be solved analytically provided the
following relation holds
\begin{equation}
L=\frac{p}{\sqrt{2K}}\ \Leftrightarrow\ A=\pi^{2}\frac{p^{2}}{K}%
\ \Leftrightarrow A=\left(  \frac{2\pi p}{ef_{\pi}}\right)  ^{2}\ ,
\label{quantuzationcondition}%
\end{equation}
where we have expressed explicitly the quantization condition in terms of the
Pions decay constant using Eqs. (\ref{action0}) and (\ref{action0.1}). Hence,
the BPS bound (to be defined here below) can be saturated when the surface
area of the layers is quantized in terms of the Pions decay constant. When the
above equation is satisfied, the solution of the Hamilton-Jacobi equation
(\ref{HJBPS1}) is%
\begin{align}
W  &  =\frac{4K^{3/2}}{pL_{r}}u\cos H\ \Rightarrow\label{HJBPS1.5}\\
\frac{\partial W}{\partial H}  &  =-\frac{4K^{3/2}}{pL_{r}}u\sin
H\ ,\ \frac{\partial W}{\partial u}=\frac{4K^{3/2}}{pL_{r}}\cos H\ .\nonumber
\end{align}
The above solution of the Hamilton-Jacobi equation associated to the present
theory allows to rewrite the energy-density in a BPS style as follows:%
\begin{equation}
T_{00}=\frac{1}{e^{2}}\left\{  \frac{K}{2\left(  pL_{r}\right)  ^{2}}\left[
\left(  pH^{\prime}\pm4K^{1/2}L_{r}u\sin H\right)  ^{2}+4\left(  u^{\prime}\mp
pK^{1/2}L_{r}\cos H\right)  ^{2}\right]  \pm\frac{dW}{dr}\right\}
\label{HJBPS2}%
\end{equation}
where $W$ in Eq. (\ref{HJBPS1.5}) is the solution the Hamilton-Jacobi equation
(\ref{HJBPS1}) (in the following we will consider the upper signs in the above
equation, the analysis with the lower signs is analogous). Consequently, it is
possible to derive the following bound:%
\begin{equation}
E=\int\sqrt{-g}d^{3}xT_{00}=AL_{r}\int_{0}^{2\pi}T_{00}dr\geq\left\vert
Q\right\vert \ ,\ \ Q=\frac{AL_{r}}{e^{2}}\left\vert W(2\pi)-W(0)\right\vert
\ , \label{BPSnew}%
\end{equation}
where the area $A$ of the Baryonic layers is quantized according to Eq.
(\ref{quantuzationcondition}).

One could naively think that such BPS trick should not work in the present
case since (in the case without Maxwell) the NLSM does not allow the
derivation of a BPS bound (and this is the reason why Skyrme introduced the
Skyrme term). In fact, as it happens in the case of the Abelian Higgs model,
the presence of the minimal coupling with the Maxwell gauge theory changes
considerably the situation. Therefore, the two main questions to answer are:

1) \textbf{Does the saturation of the bound implies the second order field
equations?}

2) \textbf{Can the first order BPS equation be actually solved?}

\textit{First of all}, the first order BPS equations%
\begin{align}
H^{\prime}+\frac{4K^{1/2}L_{r}}{p}u\sin H  &  =0\label{BPS1}\\
u^{\prime}-pK^{1/2}L_{r}\cos H  &  =0 \label{BPS2}%
\end{align}
actually imply the second order field equations in Eqs. (\ref{equ1}) and
(\ref{equ2}) as it can be verified directly by deriving Eqs. (\ref{BPS1}) and
(\ref{BPS2}) with respect to $r$ and comparing the result with Eqs.
(\ref{equ1}) and (\ref{equ2}). The above BPS system has a very clear meaning:
$u^{\prime}$ (which represent the intensity of the magnetic field) is maximal
when $\left\vert \cos H\right\vert \sim1$ which implies that $H^{\prime}\sim0$
(since $\left\vert \sin H\right\vert \sim0$), on the other hand, when
$H^{\prime}$ is large the magnetic field is small: the physical interpretation
of this fact will be discussed in a moment.

One of the most important physical implications of the present exact results
is that these allow to derive analytically the non-linear relation between the
Baryonic charge and the magnetic flux. In order to achieve this goal, instead
of solving directly Eqs. (\ref{BPS1}) and (\ref{BPS2}) it is better to use the
BPS condition to derive the analytic relation between the $SU(2)$ profile $H$
and the gauge field profile $u$ as follows:%
\begin{align}
\frac{dH}{du}  &  =-\frac{4}{p^{2}}u\tan H\ \Rightarrow\nonumber\\
H(r)  &  =\arcsin\left[  \exp\left(  -\frac{2\left(  u(r)\right)  ^{2}}{p^{2}%
}+I_{0}\right)  \right]  \ , \label{BPS3}%
\end{align}
while $I_{0}$ is an integration constant (once again one can check that if one
plugs the expression for $H(r)$ in the second order field equations together
with Eqs. (\ref{BPS1}) and (\ref{BPS2}) the second order field equations are
identically satisfied).

In conclusion, using Eq. (\ref{BPS3}), the complete set of field equations is
reduced to the simple quadrature:%
\begin{equation}
u^{\prime}=pK^{1/2}L_{r}\sqrt{1-\exp\left(  -\frac{4\left(  u(r)\right)  ^{2}%
}{p^{2}}+2I_{0}\right)  }\ . \label{BPS4}%
\end{equation}
As it will be discussed in the next sections, the integration constant $I_{0}
$ can be chosen in order to satisfy the required boundary conditions.

In conclusion, the answers to the above 2 questions are affirmative in both
cases. As it will be explained in more details in the next section, these BPS
configurations represent magnetized Baryonic layers possessing a non-trivial
Baryonic charge as well as a non-vanishing magnetic flux.\ To the best of
author`s knowledge, this is the first example of a BPS bound which can be
saturated with an explicit analytic configuration in the gauged NLSM-Maxwell
theory in (3+1) dimensions which is static, magnetized and with non-vanishing
Baryonic charge. Such a surprising result has very intriguing phenomenological
consequences on the relations between magnetic flux, Baryonic charge and
Baryonic chemical potential (which will be discussed in the following
sections). Obviously, there are many very nice papers on solitons in the
gauged NLSM Maxwell theory (see \cite{gaugsol}-\cite{gaugesky2}\ and
references therein), however, most of them employ numerical methods which
prevent, for instance, to disclose the explicit relation between the Baryonic
charge and the magnetic flux. On the more theoretical side, the fact that, in
the gauged NLSM-Maxwell theory it is possible to construct non-trivial BPS
bound which can be saturated opens an intriguing possibility related to
supersymmetry (SUSY henceforth) as well as to the findings in \cite{NittaSUSY}%
. The first obvious observation is that whenever a non-trivial BPS bound which
can be saturated is found, it is natural to ask whether such a bound is a
manifestation of some hidden SUSY. In \cite{NittaSUSY} the authors discussed a
supersymmetric extension of the Skyrme model: due to the auxiliary field
equations, such a model lacks a kinetic term. On the other hand, in the
present paper it has been shown that the gauged NLSM-Maxwell theory could
possess a non-linear realization of SUSY, which manifests itself in the fact
that the natural charge appearing in the BPS bound is a non-linear function of
the "obvious topological charge". Since the NLSM is the natural kinetic term
for Skyrme theory, one may wonder whether the inclusion of the minimal
coupling with Maxwell theory could allow a supersymmetric version of the
Skyrme model with a standard kinetic term: I hope to come back on this issue
in a future publication.

\subsection{Topological and electric currents and magnetic field}

The behavior of the magnetic field and of the $SU(2)$ profile can be very
easily described thanks to availability of the BPS equations (\ref{BPS1}),
(\ref{BPS2}) and (\ref{BPS3}). First of all, the magnetic field $\mathbf{B}%
_{i}=\varepsilon_{ijk}F_{jk}$ (whose intensity is proportional to $u^{\prime}%
$) only possesses components tangent to the Baryonic layer ($\left\vert
\mathbf{B}_{\theta}\right\vert =\left\vert \mathbf{B}_{\varphi}\right\vert
\neq0$) while the component orthogonal to the layer vanishes: $\mathbf{B}%
_{r}=0$. To see this, it is enough to remind that the energy and Baryon
densities of this non-homogeneous condensate only depend on $r$ (so that its
position can be identified with the local maximum of the Baryon density).
Moreover, from Eq. (\ref{BPS1}) and (\ref{BPS2}) one can see that the magnetic
field $u^{\prime}$ is maximal where $\left\vert \cos H\right\vert \sim1$ (so
that $\sin H\sim0$ which implies $H^{\prime}\sim0$). Consequently, the
magnetic field reaches its highest intensity where $H^{\prime}$ vanishes. This
is very interesting since the places where $H^{\prime}$ vanishes correspond to
the regions where the $SU(2)$ chiral field does not contribute to the Baryon
density (since the contribution to the Baryon density of the $SU(2)$ valued
chiral field is proportional to $H^{\prime}$). This implies that the $SU(2)$
chiral field acts as a superconductor in the sense that it tends to suppress
the magnetic field in its interior and, moreover, the magnetic field is
tangent to the chiral soliton.

In this respect, an important issue must be discussed: the magnetic flux
inside the layer should be supplemented by the condensation of the electric
charges outside. In other words, the question is: \textit{which is the
structure of layers between the magnetic-flux carrying ones?} The exact answer
to this question is very complicated at microscopic level as it involves the
analysis of quarks dynamics strongly coupled with the present BPS chiral
magnetized solitons (I will come back on this issue in a future publication).
On the other hand, at a qualitative level, the structure of \textit{layers
between the magnetic-flux carrying ones} can be described well in terms of the
gauged NLSM itself. Indeed, in order to achieve this goal, one needs exact
solutions of the gauged NLSM with the shape of a layer (so that these can fit
between the magnetic-flux carrying ones) and with non-vanishing electric
charge density (the property to be topologically stable would be extremely
welcome as well). This type of exact solutions would be very good candidates
to be the \textit{layers between the magnetic-flux carrying ones}. In the
references \cite{crystal1} it has been possible to construct analytic
solutions of the gauged NLSM and Skyrme model representing Baryonic layers
with non-vanishing electric charge density which could not be deformed
continuously to the trivial configuration due to the non-triviality of the
topological charge. These configurations in \cite{crystal1} are good
candidates to be the sought \textit{layers between the magnetic-flux carrying
ones}. The details of this interpretation will be discussed in a future publication.

Using the explicit expression of $H$ in terms of $u$ in Eq. (\ref{BPS3}) one
can compute explicitly the Baryonic charge, the new topological charge in Eq.
(\ref{BPSnew}) and the magnetic flux. This allows to answer to the following
questions: once the magnetic flux is given, which is the allowed value of the
Baryonic charge $B$ (and viceversa)? Which is the critical value of the
Baryonic chemical potential beyond which these solitons become
thermodynamically unstable? Within neutron stars as well as in Heavy Ions
Collisions \cite{HIC} \cite{QGBook}, where such non-homogeneous Baryonic
condensates are expected, such questions are of utmost importance. In order to
proceed, one observes that from the first order BPS equations one arrives at%
\begin{align}
u^{\prime}  &  =pK^{1/2}L_{r}\sqrt{1-\exp\left(  -\frac{4\left(  u(r)\right)
^{2}}{p^{2}}+2I_{0}\right)  }\Rightarrow\nonumber\\
U^{\prime}  &  =K^{1/2}L_{r}\sqrt{1-\exp\left(  -4U^{2}+2I_{0}\right)
}\ ,\ \ U=\frac{u(r)}{p}\label{BCBPS0.1}\\
dr  &  =\frac{du}{pK^{1/2}L_{r}\sqrt{1-\exp\left(  -\frac{4u^{2}}{p^{2}%
}+2I_{0}\right)  }}\Rightarrow\nonumber\\
2\pi &  =\frac{1}{pK^{1/2}L_{r}}\int_{u(0)}^{u(2\pi)}\frac{du}{\sqrt
{1-\exp\left(  -\frac{4u^{2}}{p^{2}}+2I_{0}\right)  }}. \label{BCBPS1}%
\end{align}
First of all, from Eq. (\ref{BCBPS0.1}), one can define a "rescaled" variable
$U=u/p$ in such a way that the corresponding first order BPS equation for the
"rescaled" function $U$ does not depend on $p$ at all so that, in particular,
$U$ does not depend on $p$ locally. At a qualitative level, this tells that
$u(r)$ (and consequently $u(2\pi)$) scales linearly with $p$. The above
equation (\ref{BCBPS1}) links the values of $u(r)$ at $0$ and $2\pi$ to the
integration constant $I_{0}$. Although it is not the most general option, it
is convenient to assume $u(0)=0$ since this choice (which can be achieved with
a gauge transformation) clarifies the relations between the magnetic flux, the
Baryon charge and the topological charge appearing in the BPS bound. Hence,
Eq. (\ref{BCBPS1}) becomes%
\begin{equation}
2\pi=\frac{1}{K^{1/2}L_{r}}\int_{0}^{\frac{u(2\pi)}{p}}\frac{d\tau}%
{\sqrt{1-\exp\left(  -4\tau^{2}+2I_{0}\right)  }}\ ,\ \ u(0)=0\ .
\label{BCBPS2}%
\end{equation}
In order to disclose the physical meaning of $u(2\pi)$ let us compute the
total magnetic flux $\mathbf{\Phi}$ in the $\varphi$-direction (which is
proportional to the magnetic flux along the $\theta$ direction):%
\begin{align}
\mathbf{\Phi}  &  =LL_{r}\int drd\theta F_{r\theta}=\frac{p\pi L_{r}}%
{\sqrt{2K}}\left(  u(2\pi)-u(0)\right)  \Longrightarrow\label{charge5}\\
u\left(  2\pi\right)   &  =\frac{\sqrt{2K}\mathbf{\Phi}}{p\pi L_{r}}\ .
\label{charge6}%
\end{align}
Therefore, $u(2\pi)$ is proportional to the magnetic flux along the $\varphi
-$direction. With the above choice of the sign in the BPS equations the
magnetic flux is positive while the Baryonic charge is negative (see Eq.
(\ref{charge3.1}) here below) while with the other choice one would obtain a
negative magnetic flux and a positive Baryonic charge. Since $u$ scales with
$p$, then the total magnetic flux $\mathbf{\Phi}$ scales with $p^{2}$:%
\begin{equation}
\frac{\mathbf{\Phi}}{p^{2}}\sim\Psi_{0}\ , \label{fluxarea}%
\end{equation}
where $\Psi_{0}$ plays the role\footnote{More precisely: $\Psi_{0}$ is
proportional to the elementary magnetic flux times an adimesnional number
which is fixed by requiring the physical boundary conditions described here
below.} of elementary magnetic flux (so that it does not depend on $p$). The
above result is reassuring since Eq. (\ref{fluxarea}) says that the magnetic
flux is proportional to the surface area of the hadronic layers.

The dependence of $I_{0}$ on $u(2\pi)$ can be determined by solving the
equation%
\[
2\pi=\frac{1}{K^{1/2}L_{r}}\int_{0}^{\frac{u(2\pi)}{p}}\frac{d\tau}%
{\sqrt{1-\exp\left(  -4\tau^{2}+2I_{0}\right)  }}\ ,
\]
for $I_{0}$ in terms of the other parameters (which can always be done
numerically). On the other hand, it is possible to determine the qualitative
behavior of such dependence by observing that (either for not too large values
of $u(2\pi)/p$ or for large values of $K^{1/2}L_{r}$) the integration variable
$\tau$ is small and for small $\tau$ one gets the estimate%
\begin{equation}
2\pi=\frac{1}{K^{1/2}L_{r}}\int_{0}^{\frac{u(2\pi)}{p}}\frac{d\tau}%
{\sqrt{1-\exp\left(  -4\tau^{2}+2I_{0}\right)  }}\sim\frac{1}{K^{1/2}L_{r}%
}\int_{0}^{\frac{u(2\pi)}{p}}\frac{d\tau}{\sqrt{1-\exp\left(  2I_{0}\right)
}}\sim\frac{1}{pK^{1/2}L_{r}}\frac{u(2\pi)}{\sqrt{1-\exp\left(  2I_{0}\right)
}} \label{charge0.0001}%
\end{equation}
Thus, taking into account Eq. (\ref{charge6}), one obtains
\begin{align}
\left(  1-\frac{\left(  u(2\pi)\right)  ^{2}}{\left(  2\pi pL_{r}\right)
^{2}K}\right)   &  \sim\ \exp\left(  2I_{0}\right)  \ \Rightarrow
\label{charge6.1}\\
\exp\left(  2I_{0}\right)   &  \sim\left(  1-\frac{\mathbf{\Phi}^{2}}{2\left(
p\pi L_{r}\right)  ^{4}}\right)  \label{charge6.2}%
\end{align}

Since $u(2\pi)$ fixes uniquely $I_{0}$ through Eq. (\ref{charge0.0001}), and
$u(2\pi)$ is in one-to-one correspondence with the magnetic flux through Eq.
(\ref{charge6}), the problem to fix the integration constant $I_{0}$ is
equivalent to the problem to find a suitable physical condition to fix the
total magnetic flux $\mathbf{\Phi}$. In order to find such physical condition,
let us consider the Baryonic charge which is proportional to the integral of
$\rho_{B}$ in Eq. (\ref{BCh1}) using Eq. (\ref{BPS3}):%
\begin{align}
\rho_{B}  &  =24p\frac{d}{dr}\left(  \Theta(r)\right)  \ ,\label{charge1}\\
\Theta\left(  r\right)   &  =u\left(  r\right)  \left(  \exp\left(
-\frac{4\left(  u(r)\right)  ^{2}}{p^{2}}+2I_{0}\right)  -1\right)
\ ,\label{charge2}\\
u(0)  &  =0\Rightarrow\ \Theta\left(  0\right)  =u\left(  0\right)
\ .\nonumber
\end{align}
Consequently, the Baryonic charge is%
\begin{align}
B  &  =\frac{1}{24\pi^{2}}\int_{S}\rho_{B}=4p\Theta\left(  2\pi\right)
\Rightarrow\label{charge3}\\
B  &  =4\frac{\sqrt{2K}\mathbf{\Phi}}{\pi L_{r}}\left(  \exp\left(
-\frac{8K\mathbf{\Phi}^{2}}{p^{4}\left(  \pi L_{r}\right)  ^{2}}%
+2I_{0}\right)  -1\right)  \ . \label{charge3.1}%
\end{align}
Hence, the integration constant $I_{0}$ must be fixed in such a way that $B$
is an integer:
\begin{equation}
\left\vert B\right\vert =4\frac{\sqrt{2K}\mathbf{\Phi}}{\pi L_{r}}\left(
1-\exp\left(  -\frac{8K\mathbf{\Phi}^{2}}{p^{4}\left(  \pi L_{r}\right)  ^{2}%
}+2I_{0}\right)  \right)  =n\ ,\ n\in%
\mathbb{N}
_{+}\ . \label{charge3.2}%
\end{equation}
The above equation is a trascendental equation for $I_{0}$ which cannot be
solved in closed form. Nevertheless, the qualitative behavior can be described
as follows. For large magnetic fluxes, the exponential terms is small so that
$B$ and $\mathbf{\Phi}$ are proportional and both of them are quantized in
terms of $n$ on the right hand side of Eq. (\ref{charge3.2}). Moreover, since
the magnetic flux is of order of $p^{2}$, the integer $n$ on the right hand
side of Eq. (\ref{charge3.2}) must be of order of $p^{2}$:%
\[
n\sim p^{2}\ .
\]
Therefore, for large values of the magnetic flux, the Baryonic charge also
grows with $p^{2}$.

In this limit in which Eqs. (\ref{fluxarea}) and (\ref{charge6.2}) are valid,
one can rewrite Eq. (\ref{charge3.2}) as a trascendental equation for the
elementary flux $\Psi_{0}$:
\begin{align}
\left\vert B\right\vert  &  =p^{2}=4p^{2}\frac{\sqrt{2K}\Psi_{0}}{\pi L_{r}%
}\left(  1-\left(  1-\frac{\Psi_{0}^{2}}{2\left(  \pi L_{r}\right)  ^{4}%
}\right)  \exp\left(  -\frac{8K\Psi_{0}^{2}}{\left(  \pi L_{r}\right)  ^{2}%
}\right)  \right)  \ \Rightarrow\nonumber\\
1  &  \approx4\frac{\sqrt{2K}\Psi_{0}}{\pi L_{r}}\left(  1-\left(
1-\frac{\Psi_{0}^{2}}{2\left(  \pi L_{r}\right)  ^{4}}\right)  \exp\left(
-\frac{8K\Psi_{0}^{2}}{\left(  \pi L_{r}\right)  ^{2}}\right)  \right)
\ \Rightarrow\label{charge3.2b}%
\end{align}%
\[
\frac{\pi L_{r}}{4\sqrt{2K}\Psi_{0}}\approx\left(  1-\left(  1-\frac{\Psi
_{0}^{2}}{2\left(  \pi L_{r}\right)  ^{4}}\right)  \exp\left(  -\frac
{8K\Psi_{0}^{2}}{\left(  \pi L_{r}\right)  ^{2}}\right)  \right)  \ .
\]
When the magnetic flux is not large numerical methods must be used. On the
other hand, in the applications in neutron stars and heavy ions collisions
(see \cite{HIC} \cite{QGBook}) one expects large values of the magnetic flux,
which is the situation which will be mostly analyzed in the following.

It is interesting to compare the Baryonic charge with the topological charge
$Q$ appearing in the BPS bound:%
\begin{align*}
Q  &  =\frac{AL_{r}}{e^{2}}\left[  W(2\pi)-W(0)\right]
\ ,\ \ u(0)=0\Rightarrow W(0)=0\Rightarrow\\
Q  &  =\frac{4\pi^{2}pK^{1/2}}{e^{2}}u(2\pi)\sqrt{1-\exp\left(  -\frac
{4\left(  u(2\pi)\right)  ^{2}}{p^{2}}+2I_{0}\right)  }\Rightarrow\\
Q  &  =4\sqrt{2}\pi\frac{K}{e^{2}L_{r}}\mathbf{\Phi}\sqrt{1-\exp\left(
-\frac{8K\mathbf{\Phi}^{2}}{p^{4}\left(  \pi L_{r}\right)  ^{2}}%
+2I_{0}\right)  }%
\end{align*}

It is easy to see that the ratio $Q/B$ of the topological charge which appears
naturally in the BPS bound over the Baryon charge is not constant:%
\begin{equation}
\left\vert \frac{Q}{B}\right\vert =\frac{\pi^{2}\sqrt{K}}{e^{2}}\frac{1}%
{\sqrt{1-\exp\left(  -\frac{8K\mathbf{\Phi}^{2}}{p^{4}\left(  \pi
L_{r}\right)  ^{2}}+2I_{0}\right)  }}\ . \label{comparison}%
\end{equation}
Therefore, when the magnetic flux is large, the two topological charges are
proportional. On the other hand, when the magnetic flux is small, the above
ratio is very large and the topological charge which appears in the bound is
much larger than the Baryonic charge. An important conclusion from the above
relation is that it may happen that the topological charge suitable to derive
a saturable BPS bound is a non-linear function of the "obvious" BPS charge
(the Baryonic charge in the present case) which may not be suitable to achieve
a BPS bound which can be saturated.

\section{Thermodynamics}

The direct computations of the classical grand canonical partition function
and the corresponding free energy of this family of BPS Magnetized Baryonic
layers presents some difficulties: it cannot be computed in a closed form due
to the fact that the dependence on the discrete labels (such as $p$ in Eq.
(\ref{ansatzmagnetico1})) of the energy of these BPS configurations is
determined by trascendental equations. The best method to analyze the above
partition function is the saddle point analysis with the refined technique of
resurgence (for detailed reviews see \cite{resurgence1} \cite{resurgence2}
\cite{resurgence3} and references therein). I hope to come back soon on the
analysis of the resurgent behavior of the partition function in Eq.
(\ref{thermo2}). On the other hand, in the limit of large magnetic flux (when
Eqs. (\ref{fluxarea}) and (\ref{charge6.2}) are valid) one can use the
proportionality of the magnetic flux with $p^{2}$: $\mathbf{\Phi}\approx
\Psi_{0}p^{2}$ to estimate the critical Baryonic chemical potential $\mu
^{\ast}$ beyond which one expects a change in the partition function. As it
has been already discussed, this is an excellent approximation in many
situations of physical interest (such as neutron stars and Heavy Ions
Collisions \cite{HIC} \cite{QGBook}). The classical grand canonical partition
function is the sum over the discrete label $p$ of the factor $\exp\left[
-\beta\left(  E\left(  p\right)  +\mathbb{P}V-\mu_{B}B\right)  \right]  $
(where $\beta$ is the inverse temperature, $\mathbb{P}$ is the pressure, $V$
is the volume and $\mu_{B}$ is the Baryonic chemical potential and $E(p)$ is
in Eq. (\ref{BPSnew})). Both the pressure and the volume (due to the
quantization condition in Eq. (\ref{quantuzationcondition})) depend on the
discrete integer label $p$:%
\[
V=V(p)=2\pi^{3}L_{r}\frac{p^{2}}{K}%
\]%
\begin{align*}
\mathbb{P}  &  \mathbb{=P(}p\mathbb{)=-}\frac{\partial E\left(  p\right)
}{\partial V}=\mathbb{-}\frac{\partial L_{r}}{\partial V}\frac{\partial
E\left(  p\right)  }{\partial L_{r}}=-\frac{K}{2\pi^{3}p^{2}}\frac{\partial
E\left(  p\right)  }{\partial L_{r}}\Rightarrow\\
\mathbb{P}V  &  =-L_{r}\frac{\partial E\left(  p\right)  }{\partial L_{r}}\ ,
\end{align*}
where, in the above formulas, we have used the fact that for fixed $p$ the
volume is proportional to $L_{r}$. It is worth to note here that the product
$\mu_{B}B\left(  p\right)  $ should be considered as positive with both
choices of the sign in the BPS equations. Indeed, when $B$ is negative (which
is the case when the flux is positive, as it has been already mentioned)
$\mu_{B}$ also must be negative (since the chemical potential of the
anti-Baryons is opposite to the chemical potential of the Baryon which is
positive: see \cite{R11} and references therein). Hence, in this
approximation, the grand canonical partition function $Z$ reads:
\begin{align}
Z  &  =\sum_{p=-\infty\ ,\ p\neq0}^{+\infty}\exp\left[  -\beta\left(  E\left(
p\right)  -L_{r}\frac{\partial E\left(  p\right)  }{\partial L_{r}}-\mu
_{B}B\left(  p\right)  \right)  \right]  \ =Z\left(  \beta,\mu_{B}%
,L_{r}\right) \label{thermo2}\\
E\left(  p\right)   &  =4\sqrt{2}\pi\frac{K}{e^{2}L_{r}}\left\vert \Psi
_{0}\right\vert p^{2}\sqrt{1-\exp\left(  -\frac{8K\Psi_{0}^{2}}{\left(  \pi
L_{r}\right)  ^{2}}+2I_{0}\right)  }\ ,\nonumber\\
\mu_{B}B\left(  p\right)   &  =\mu_{B}p^{2}\ ,\ \ \mu_{B}\geq0\ .\nonumber
\end{align}

Consequently, in this limit one can express the partition function $Z$ in
terms of the elliptic theta function%
\begin{align}
Z  &  =\sum_{p\in%
\mathbb{Z}
}\left(  q^{1/2}\right)  ^{p^{2}}-1\ ,\ \ q^{1/2}=\exp\left[  -\beta\left(
a\left(  \Psi_{0}\right)  -L_{r}\frac{\partial a\left(  \Psi_{0}\right)
}{\partial L_{r}}-\mu_{B}\right)  \right]  \ ,\label{thermo3}\\
a\left(  \Psi_{0}\right)   &  =4\sqrt{2}\pi\frac{K}{e^{2}L_{r}}\left\vert
\Psi_{0}\right\vert \sqrt{1-\exp\left(  -\frac{8K\Psi_{0}^{2}}{\left(  \pi
L_{r}\right)  ^{2}}+2I_{0}\right)  }\approx2^{7/4}\pi^{3/2}\frac{K^{3/4}%
}{e^{2}}\left\vert \frac{\Psi_{0}}{L_{r}}\right\vert ^{1/2} \label{thermo3.11}%
\end{align}
where $K$ (which is a "rescaled" version of the Pions decay constant) is
defined in Eqs. (\ref{action0}) and (\ref{action0.1}) while Eq.
(\ref{charge3.2b}) has been taken into account and the term "-1" in Eq.
(\ref{thermo3}) is due to the fact that the $p=0$ terms is absent in the
present family since, when $p=0$ the topological charge vanishes. Thus, one
expects a change in the behavior of the grand canonical partition function
when the chemical potential $\mu_{B}$ becomes of the same order of the term
$a\left(  \Psi_{0}\right)  -L_{r}\frac{\partial a\left(  \Psi_{0}\right)
}{\partial L_{r}}$:%
\[
\mu_{B}\sim a\left(  \Psi_{0}\right)  -L_{r}\frac{\partial a\left(  \Psi
_{0}\right)  }{\partial L_{r}}\
\]
which (using Eq. (\ref{thermo3.11})) gives the following estimate \ \ \ \ \
\begin{equation}
\mu^{\ast}\ \approx3\left(  2^{3/4}\pi^{3/2}\frac{K^{3/4}}{e^{2}}\right)
\left\vert \frac{\Psi_{0}}{L_{r}}\right\vert ^{1/2}=3\left(  2^{3/4}\pi
^{3/2}\right)  \left(  \frac{f_{\pi}}{2}\right)  ^{3/2}\left\vert \frac
{\Psi_{0}}{eL_{r}}\right\vert ^{1/2}\ , \label{thermo4}%
\end{equation}
where we have expressed explicitly the critical chemical potential in terms of
the Pions decay constant using Eqs. (\ref{action0}) and (\ref{action0.1}).
Obviously, a more careful analysis of the partition function is needed as, for
instance, in order to justify the fact that the chemical potential for
configurations with negative $B$ is the opposite of the chemical potential of
the configurations with positive $B$ a proper semi-classical treatment of
these solitons would be necessary\footnote{It is worth emphasizing that, in
systems with dominant magnetic interactions, the fact that the chemical
potential of the particles is opposite to the chemical potential of the
anti-particles extends to the world of quasi-particles as well. For instance,
in \cite{chemicalPotential} it has been shown that the two magnon eigenmodes
of the system are described by an equal and opposite chemical potential, in
analogy with a particle-antiparticle pair.}. Nevertheless, the above estimate
in Eq. (\ref{thermo4}) provides with the correct qualitative behavior in the
classical limit which is not spoiled in the regime in which the semi-classical
corrections are small. Therefore, the critical Baryonic chemical potential
grows with $\left\vert \frac{\Psi_{0}}{L_{r}}\right\vert ^{1/2}$. This means,
in a sense, that the bigger is $\left\vert \frac{\Psi_{0}}{L_{r}}\right\vert
$, the more difficult is to destabilize these hadronic layers. The above
results also imply that in order for these configurations to survive in a
large box (large $L_{r}$), $\Psi_{0}$\ must increase accordingly. In view of
the difficulties that LQCD faces when dealing with Baryonic chemical
potential, the present formalism can prove very useful for phenomenological
applications at finite Baryon density with non-vanishing magnetic fluxes. I
will come back on these relevant issue in a future publication.

\section{Conclusions and perspectives}

Using the equation of Hamilton-Jacobi from classical mechanics, in the present
paper the first analytic examples of static non-homogeneous condensates
possessing both Baryonic charge and magnetic flux have been constructed. Such
configurations saturate a novel BPS bound (provided the surface area of the
Baryonic layer is quantized) in which the topological charge $Q$ is a
non-linear function of the Baryonic charge (which is the "obvious" topological
charge which one would consider at first in a BPS bound). The present
formalism allows to compute the Baryonic charge as a function of the magnetic
flux as well as the dependence of $Q$ on the Baryonic charge. The computation
of the classical grand canonical partition function associated to this family
of BPS configurations allows to estimate the Baryonic chemical potential
beyond which such configurations become thermodynamically unstable. The
present formalism can have a huge impact on the analysis of chiral
perturbation theory and, more generically, of QCD under extreme conditions of
non-vanishing Baryon density, strong magnetic fields and low temperatures
(where LQCD experiences problems and analytic perturbative methods are
ineffective). For instance, it allows to determine how much magnetic flux is
needed to increase the Baryonic charge of the condensate (a computation which
would be difficult to do with other methods). Another interesting application
of the present results is the analysis of the coupling of the quarks to such
gauged magnetized BPS Baryonic configurations (using the well known Dirac-like
equation which describes the interactions of quarks with $SU(2)$ chiral
fields: see \cite{quarkskyrme1} \cite{quarkskyrme2} \cite{quarkskyrme3}
\cite{quarkskyrme4}\ and references therein). I hope to come back soon on
these interesting topics in a future publication.

\subsection*{Acknowledgements}

The present author would like to warmly thank Julio Oliva (who participated in
the early stages of this projects) and Marcelo Oyarzo for illuminating
discussions and suggestions. The present author would also like to thank the
anonymous referee especially for the suggestions on future works. This work
has been funded by Fondecyt grants 1200022\textbf{. }The Centro de Estudios
Cientificos (CECs) is funded by the Chilean Government through the Centers of
Excellence Base Financing Program of Conicyt.

\end{document}